\documentclass{ptapap}
\usepackage{hyperref}
\author{T. P. Adhikari}[CAMK]
\author{A. R\'o\.za\'nska}[CAMK]
\author {K. Hryniewicz}[CAMK]
\author {B. Czerny}[CAMK,CFT]
\affil[CAMK]{Nicolaus Copernicus Astronomical Center, Polish Academy of Sciences,\\
  Bartycka 18, 00--716 Warsaw, Poland}
\affil[CFT]{Centre for Theoretical Physics,Polish Academy of Sciences,\\
Al. Lotnikow 32/46 02-668, Warsaw, Poland}

\title{Absorption Measure Distribution in AGN}

\begin{document}

\maketitle

\begin{abstract}

The most common observed feature of the X-ray outflows in 
active galactic nuclei (AGN) is their broad ionization 
distribution spanning up to $\sim$ 5 orders of magnitude in ionization levels. 
This feature is quantified in terms of absorption measure distribution (AMD), 
defined as the distribution of column density with the ionization parameter. 
Recently, the photoionization models with constant pressure assumption 
are shown to well reproduce the observed shape of AMD. However, there exist 
inconsistencies in the normalization and the position of discontinuities presented 
in the AMD shape between the observation and model. 
In this work, we show that AMD normalization differs by one order of magnitude 
depending on the shape of ionizing spectral energy distribution (SED).

\end{abstract}

\section{Absorption Measure Distribution}

One of the major merit of high resolution X-ray spectroscopy was the discovery of
highly ionized absorbing regions in the 50\% of AGN commonly named warm absorbers (WA).  
The detection of X-ray blue shifted absorption, in particular different states of iron, has 
shown that the warm absorber is broadly stratified in 
ionization levels. The ionic column density for the most abundant 
 individual ions is obtained by fitting the observed high resolution X-ray lines. 
Those ionic column densities can be related to total column density of the gas 
only by adopting photoionization model. 
 
Absorption measure distribution (AMD)in AGN outflows is defined as the distribution of matter column density with 
the ionization parameter along the line of sight, i.e., 
AMD = $dN_{\rm H} /d (\log \xi$) \citep{holczer2007} where, $N_{\rm H}$ is the total 
column density  and $\xi$ is the ionization parameter. 
For the detail procedure of derivation of AMD from observations, 
we refer the reader to the paper 
by \citet{holczer2007}, but we would like to point out that any step of analysis 
always requires the high resolution X-ray data. Therefore, up to now we know 
only 7 objects with AMD determined from observations \citep{behar2009,detmers2011,stern2014}.

Observations indicated that in all studied Sy galaxies, derived AMDs span up to 
5 orders of magnitude in $\xi$ with the same normalization located at the level of  $N_{\rm H} \sim 10^{22}$ 
cm$^{-2}$ or slightly below  \citep{behar2009,stern2014}. Furthermore, all observed AMDs have shown 
one prominent discontinuity between log$\xi$ $\sim$ 0.5 and 1.5
\citep{stern2014, laha2014}. However, in case of Seyfert 1 galaxy Mrk 509, \citet{detmers2011}  
have shown that the AMD contains two strong discontinuities around the ionization 
degree of: log$\xi$ $\sim$  2-3 and 3-4. The common interpretation of these discontinuities 
is the presence of thermally unstable zones in the outflow, which was proved  in case of Mrk~509 with 
the use of photoionization calculations \citep{adhikari2015}. The authors have shown that 
two unstable zones are clearly present in the computed models and their position 
agrees with the observed AMD in Mrk~509. Nevertheless, the normalization of the best fitted AMD 
model  was one order of magnitude higher than the unique value obtained from observations. 

In this paper we show that the shape of the ionized spectral energy distribution (SED)
changes the AMD normalization by one order of magnitude. We have investigated systematic 
studies how AMD shape depends on physical parameters of ionized gas. All photoionization 
calculations are done with assumption that gas clouds are 
under total constant pressure  \citep{rozanska2004,rozanska2006,stern2014,adhikari2015,goosmann2016}.
One of the most important consequence of the constant
pressure models is self consistent stratification in the density and 
ionization across a single gas clouds. 
\citet{stern2014} has shown that the observed AMD normalization can 
be well produced by radiation pressure confinement (RPC)
model, i.e. constant pressure model in {\sc cloudy} \citep{ferland2013} 
numerical code. Nevertheless, RPC model computed by {sc cloudy} does not explain 
prominent discontinuity presented in the observed AMD. For our research we use 
photoionization code {\sc titan} \citep{dumont2000}, which can well reproduced observed 
discontinuities caused by thermal instability \citep{adhikari2015}.

Among many computed models, in this paper we present two extreme cases of SEDs.
Considered spectral shapes  differ in the strength  of optical/UV disk emission and 
the ratio of hard X-ray power law luminosity to the luminosity of an accretion disk. 
We show that AMD normalization agrees with the observed one, when illuminated SED 
displays strong disk component with luminosity 100 times larger than X-ray power law luminosity.

\section{Photoionization models}

\begin{figure}[!h]
\centering
\includegraphics[width=6.3cm]{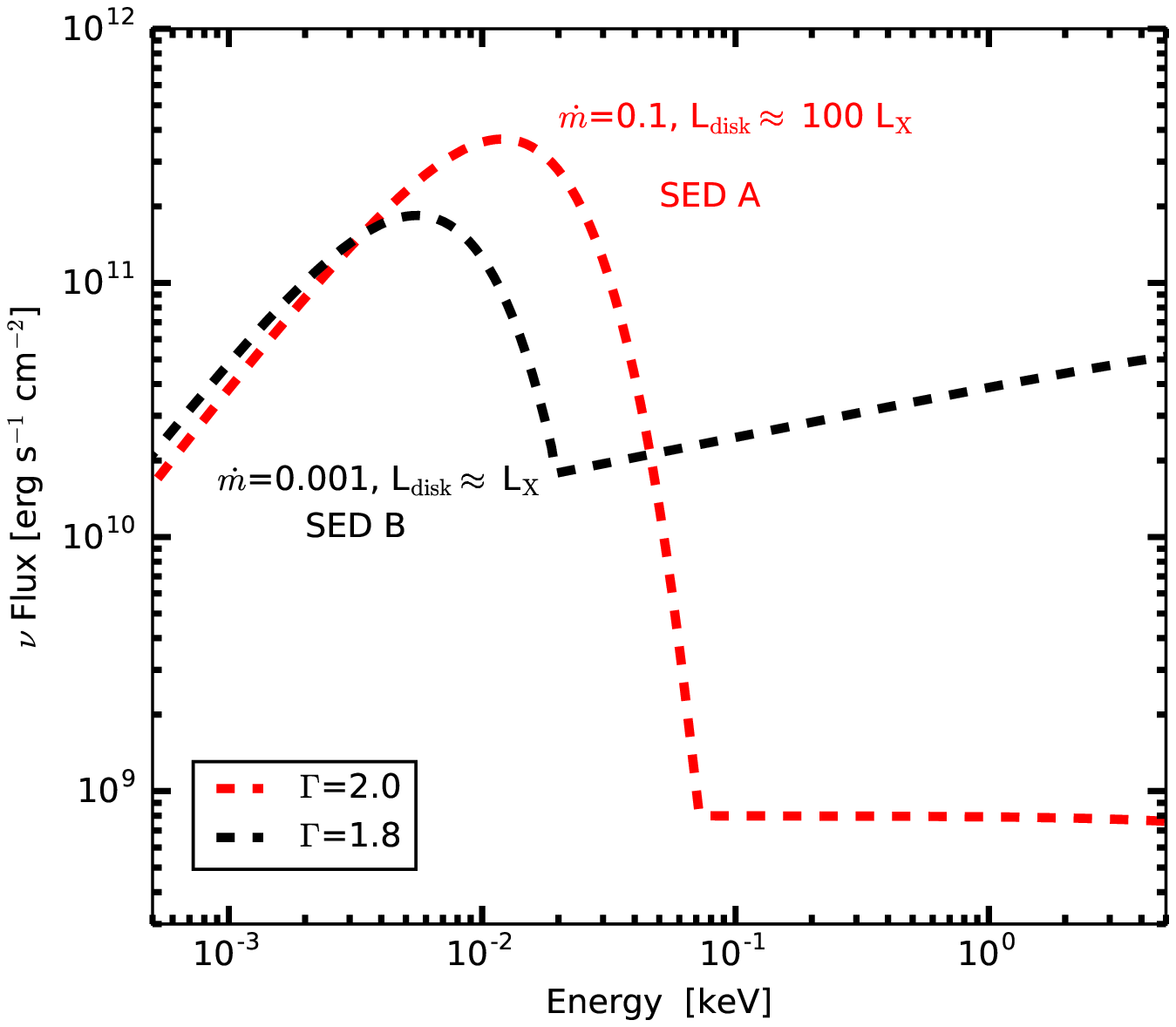}
\includegraphics[width=6.3cm]{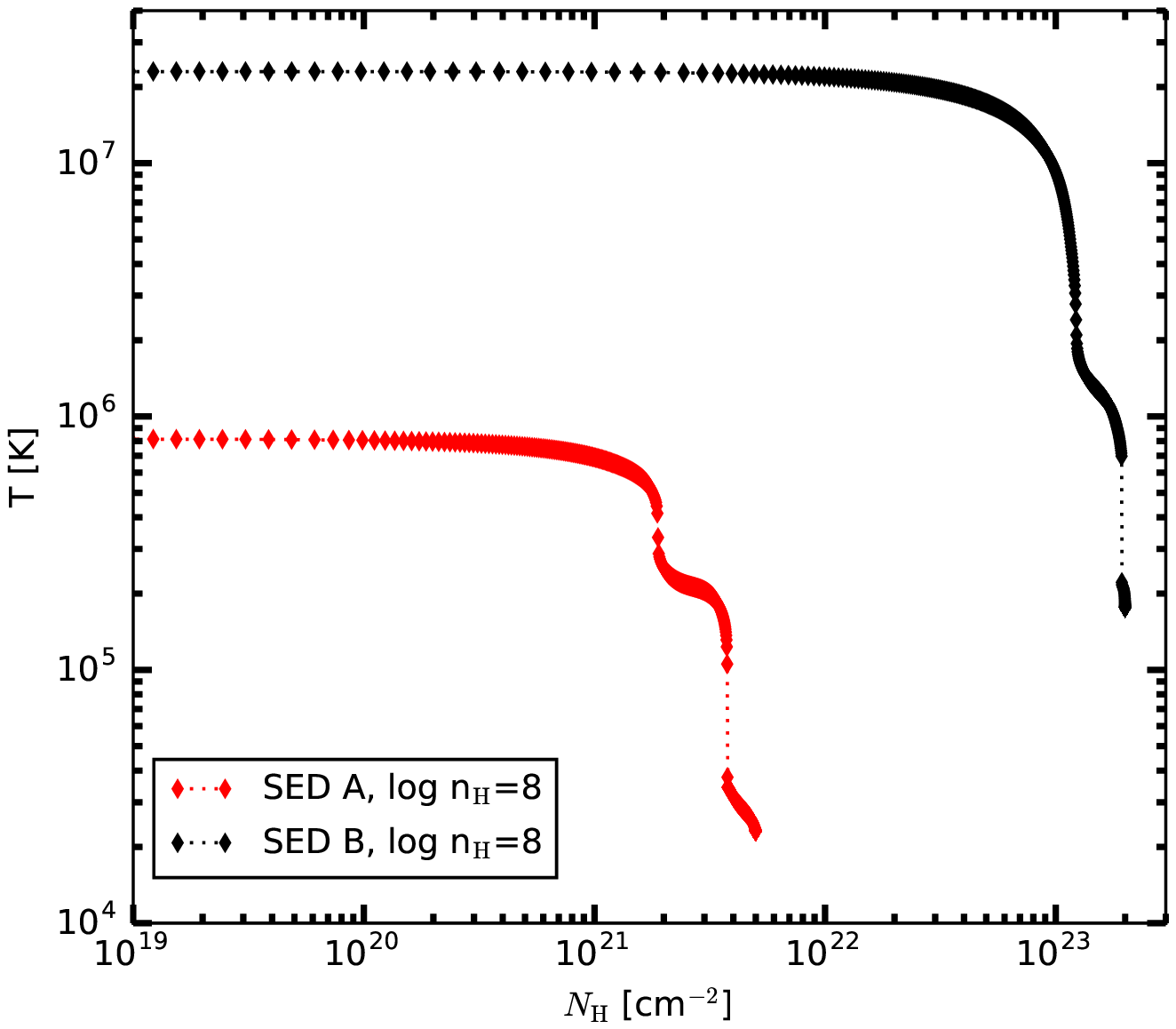}
\caption{Two SED shapes used in this paper are presented in left panel, while right panel represents 
temperature structure across WA for the gas illuminated by those SEDs: SED A 
and SED B.}
\label{fig:sed}
\end{figure}

We compare here two SED shapes obtained by varying the mass accretion rates $\dot{m}$
and the ratio of disk luminosity to the X-ray power law luminosity $L_{\rm disk}/L_{\rm X}$.
For the computations of the disk contribution to the SED, we took the multi-color black body emission 
from  the accretion disk around black hole of 
mass $M_{\rm BH}$ = 10$^{8}$ M$_{\odot}$ with two values of accretion rate which differ by 
the factor of 100. Hard X-ray emission is assumed to be a power law with photon index 
$\Gamma$ chosen in the way to fulfill relative normalization required by us. 
The final SED shapes, named SED A and SED B are obtained by taking i) $\dot{m}$= 0.1 
and $L_{\rm disk}/L_{\rm X}$= 100 and ii) $\dot{m}$= 0.001 and $L_{\rm disk}/L_{\rm X}$= 1
respectively. The resulting SED shapes are shown in Fig.~\ref{fig:sed} left panel.

We performed photoionization calculations using the numerical code {\sc titan}
assuming that ionized gas is in total pressure equilibrium i.e. 
$P_{\rm gas} + P_{\rm rad}$=const at each optical depth of the cloud.  
For the detail properties of {\sc titan} code and its capability to solve the 
radiative transfer accounting all the appropriate physical processes,
we refer the reader to the relevant literature \citep{collin2004,goncalves2006,rozanska2006}.
We assumed plane parallel slab of gas and assign the gas density at its surface to 
be 10$^{8}$ cm$^{-3}$, which increases in the inner layers due to the 
compression by incident radiation field. As the result, WA is strongly stratified, and the 
appropriate temperature gradients are presented in  Fig.~\ref{fig:sed} right panel. 
Geometrically thick, hot heated layer is accompanied by strong temperature gradient on the 
back side of the cloud, where ionization drops steeply. The resulting ionization front is geometrically very thin. 

\begin{figure}[!h]
\centering
\includegraphics[width=6.3cm]{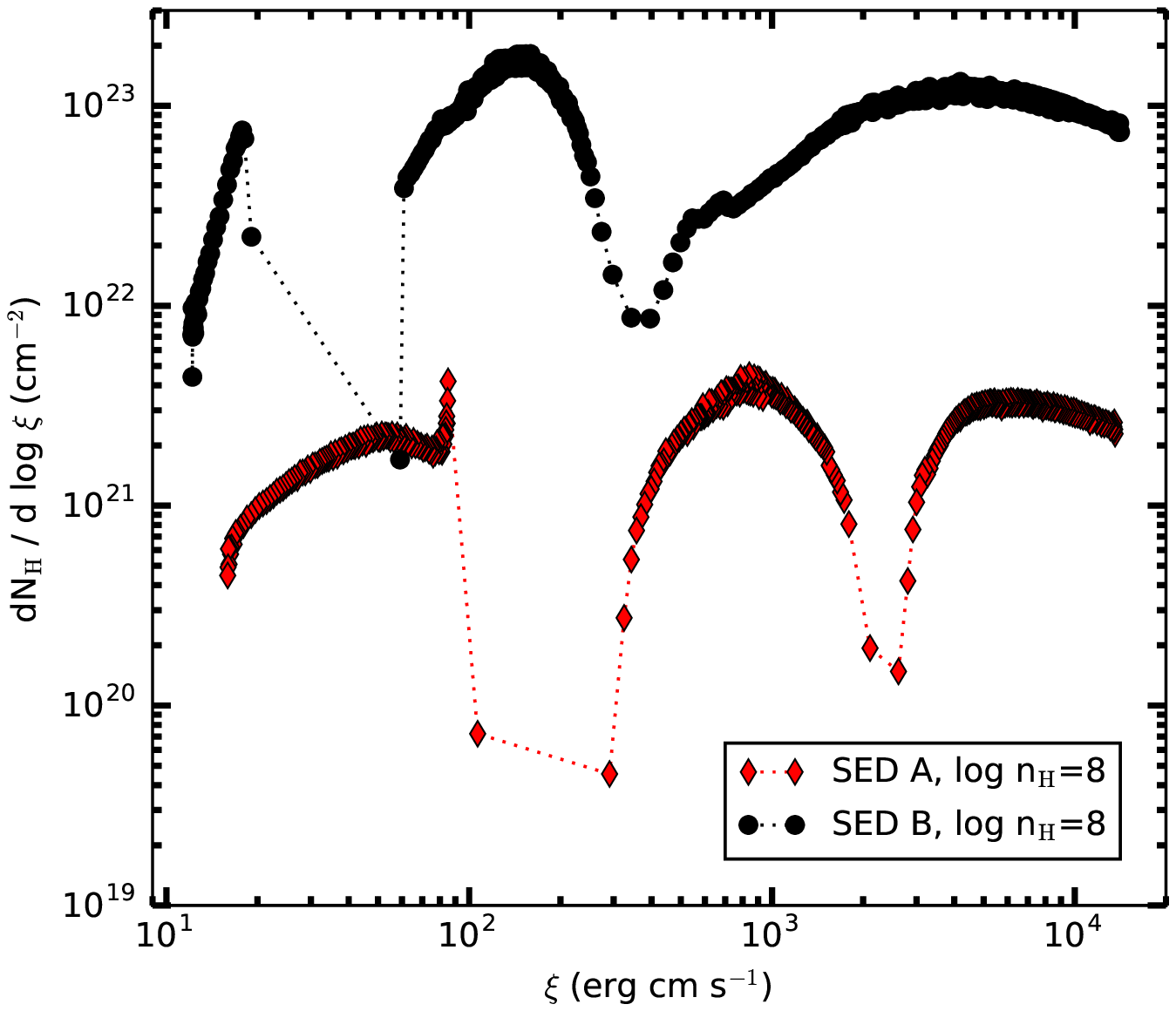}
\includegraphics[width=6.3cm]{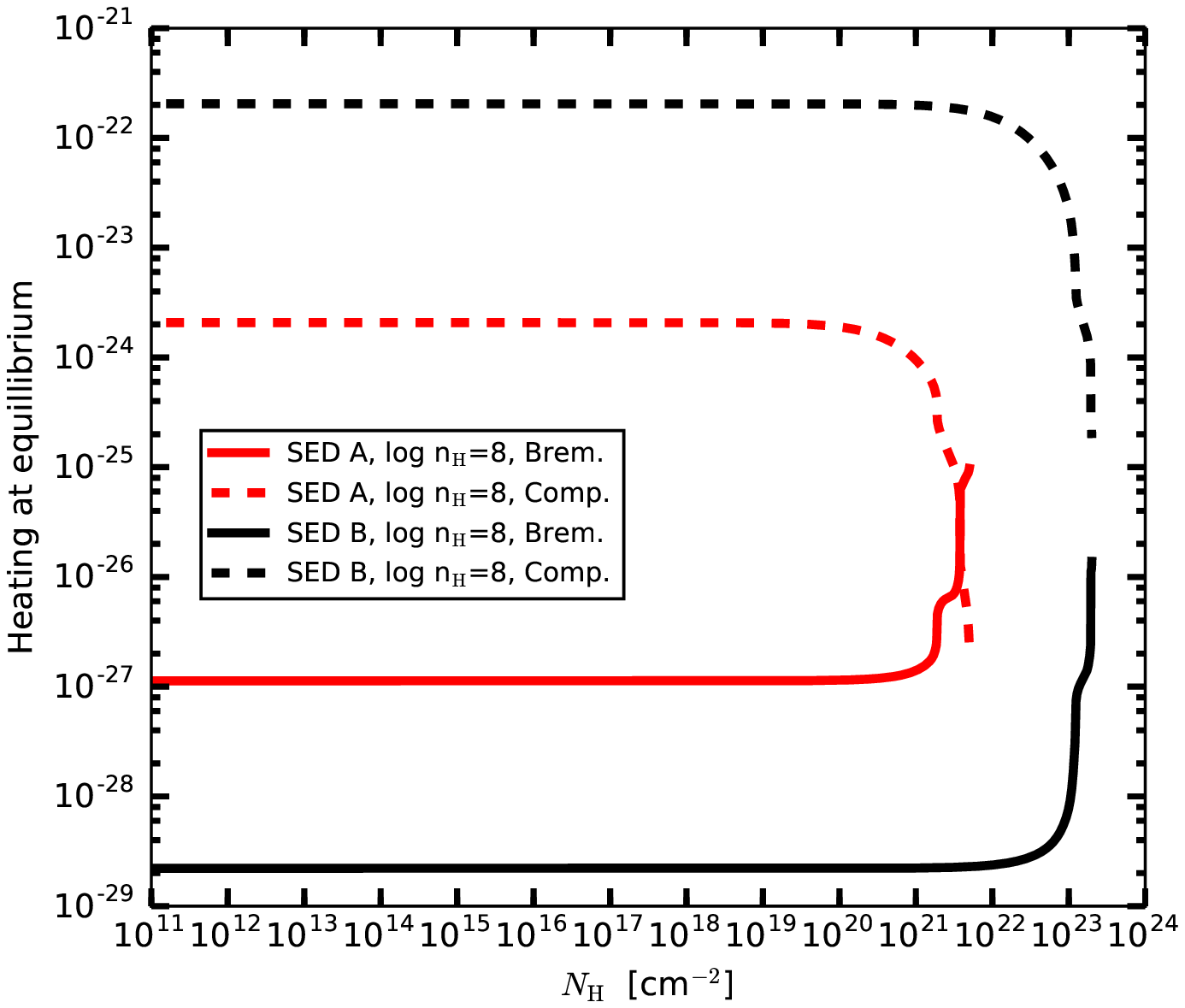}
\caption{AMD models for two spectral shapes: SED A (red) and SED B (black) for
the gas density 10$^8$ cm$^{-3}$ at the cloud surface. Right panel shows the amount of heating 
caused by Compton and Bremsstrahlung processes across both zones.}
\label{fig:amd}
\end{figure}

\section{Results and discussion}
The AMD models computed with {\sc titan}
photoionization code are shown in the right panel of Fig.~\ref{fig:amd}.
We found that the AMD models obtained for the SED B has higher normalization
by the factor of $\sim$25 than the models with the SED A. These results 
show that the shape of the incident radiation changes 
the normalization of AMD models. Strong disk component with weak
hard X-ray power law produces lower AMD normalization in agreement with observations. 

\citet{adhikari2015} have found that the AMD structure depends on the 
gas density. Our conclusion is that the dependence of AMD on the SED and the gas
density is degenerated and complex. Final conclusion from our model computations for
large parameters space  is in progress and will be reported in the future paper.
The next generation X-ray mission {\it ATHENA} with its high resolution 
instrument {\it X-IFU}, will be able to resolve more 
absorption lines with unprecedented details allowing for better 
understanding of AMD nature from observations.

\acknowledgements{This research was supported by Polish National 
Science Center grants No.2015/17/B/ST9/03422, 2015/18/M/ST9/00541, and 
2016/21/N/ST9/03311.}

\bibliographystyle{ptapap}
\bibliography{ptapapdoc}

\end{document}